%% file: root.tex
\documentclass[letterpaper, 10 pt, conference]{ieeeconf}
\IEEEoverridecommandlockouts

\usepackage{amsmath, amsthm, amssymb, amsfonts, mathtools, xargs, tensor, units, cite, esint, stmaryrd, mathrsfs, color, algpseudocode, graphicx}
\usepackage[unicode=true, bookmarks=true,bookmarksnumbered=false,bookmarksopen=false,breaklinks=false,pdfborder={0 0 1},backref=false,colorlinks=false]{hyperref}

\title{Inverse reinforcement learning in continuous time and space}

\author{Rushikesh Kamalapurkar\thanks{Rushikesh Kamalapurkar is with the School of Mechanical and Aerospace Engineering at the Oklahoma State University. Email: \texttt{rushikesh.kamalapurkar@okstate.edu}}}

\input{Macros.tex}

\input{Theorems.tex}

\begin{document}
\maketitle
\begin{abstract}
	This paper develops a data-driven inverse reinforcement learning technique for a class of linear systems to estimate the cost function of an agent online, using input-output measurements. A simultaneous state and parameter estimator is utilized to facilitate output-feedback inverse reinforcement learning, and cost function estimation is achieved up to multiplication by a constant.
\end{abstract}
\section{Introduction}
Seamless cooperation between humans and autonomous agents is a vital yet challenging aspect of modern robotic systems. Effective cooperation between humans and autonomous systems can be achieved if the autonomous systems are capable of learning to act by observing other cognitive entities. Based on the premise that a cost (or reward) function fully characterizes the intent of the demonstrator, a method to learn the cost function from observations is developed in this paper. The cost-estimation problem first appears in \cite{SCC.Kalman1964} in a linear-quadratic regulation (LQR) setting, and a solution is provided in \cite{SCC.Boyd.Ghaoui.ea1994} via linear matrix inequalities. For nonlinear systems and cost functions, computation of closed-form solutions is generally intractable, and hence, approximate solutions are sought.

In \cite{SCC.Ng.Russell2000,SCC.Abbeel.Ng2004,SCC.Abbeel.Ng2005,SCC.Ratliff.Bagnell.ea2006}, the cost function of a Markov decision process (MDP) is learned using inverse reinforcement learning (IRL). It is demonstrated that the IRL problem is inherently ill-posed in the sense that it has multiple possible solutions, including the trivial ones. To overcome the degeneracy, the cost function that differentiates the optimal behavior from the suboptimal behaviors \textit{by a margin} is sought. In \cite{SCC.Ziebart.Maas.ea2008} the maximum entropy principle (cf. \cite{SCC.Jaynes1957}) is utilized to solve the ill-posed IRL problem for deterministic MDPs. In \cite{SCC.Ziebart.Bagnell.ea2010} a causal version of the maximum entropy principle is developed and utilized to solve IRL problems in a fully stochastic setting. An IRL algorithm based on minimization of the Kullback-Leibler divergence between the empirical distribution of trajectories obtained from a baseline policy and the trajectories obtained from the cost-based policy is developed in \cite{SCC.Boularias.Kober.ea2011}.

In the past two decades, Bayesian \cite{SCC.Ramachandran.Amir2007}, natural gradient \cite{SCC.Neu.Szepesvari2007}, game theoretic \cite{SCC.Syed.Schapire2008}, and feature construction based methods \cite{SCC.Levine.Popovic.ea2010} have also been developed for IRL. IRL is extended to problems with locally optimal demonstrations in \cite{SCC.Levine.Koltun2012} using likelihood optimization and to problems with nonlinear cost functions in \cite{SCC.Levine.Popovic.ea2011} using Gaussian processes (GP). Another GP-based IRL algorithm that increases the efficiency and applicability of IRL techniques by autonomously segmenting the overall task into sub-goals is developed in \cite{SCC.Michini.How2012}. Over the years, intent-based approaches such as IRL have been successfully utilized to teach UxVs and humanoid robots to perform specific maneuvers in an \textit{offline} setting \cite{SCC.Abbeel.Ng2004,SCC.Mombaur.Truong.ea2010,SCC.Michini.Walsh.ea2015}. 

Offline approaches are ill suited for applications where teams of autonomous agents with varying levels of autonomy work together to achieve evolving tasks. For example, consider a fleet of unmanned air vehicles where only a few of the vehicles are remotely controlled by human operators and the rest are fully autonomous and capable of synthesizing their own control policies based on the task. If the tasks are subject to change and are known only to the human operators, the autonomous agents need the ability to identify the changing objectives from observations in real-time.

 Motivated by recent progress in real-time reinforcement learning (see, e.g., \cite{SCC.Vamvoudakis.Lewis2010,SCC.Bian.Jiang.ea2014,SCC.Modares.Lewis2014,SCC.Kamalapurkar.Walters.ea2016,SCC.Wang.Liu.ea2016}), this paper develops an output-feedback IRL technique for a class of linear systems to estimate the cost function online using input-output measurements. The paper is organized as follows. Section \ref{sec:Notation} details the notation used throughout the paper. Section \ref{sec:Problem} formulates the problem. Section \ref{sec:SYSID} details the development of a simultaneous state and parameter estimator that facilitates output-feedback cost estimation. Section \ref{sec:IBE} formulates the error signal that is utilized in Section \ref{sec:IRL} to achieve online IRL. Section \ref{sec:Purge} details the purging algorithm used to facilitate IRL in conjunction with the state and parameter estimator developed in Section \ref{sec:SYSID}. Section \ref{sec:Ana} analyzes the convergence of the developed algorithm and Section \ref{sec:Con} concludes the paper.
\section{Notation}\label{sec:Notation}
The $n-$dimensional Euclidean space is denoted by $\R^{n}$. Elements of $\R^{n}$ are interpreted as column vectors and $\left(\cdot\right)^{T}$ denotes the vector transpose operator. The set of positive integers excluding 0 is denoted by $\n$. For $a\in\R,$ $\R_{\geq a}$ denotes the interval $\left[a,\infty\right)$ and $\R_{>a}$ denotes the interval $\left(a,\infty\right)$. Unless otherwise specified, an interval is assumed to be right-open. If $a\in\R^{m}$ and $b\in\R^{n}$ then $\left[a;b\right]$ denotes the concatenated vector $\begin{bmatrix}a\\
b
\end{bmatrix}\in\R^{m+n}$. The notations and $\id_{n}$ and $0_{n}$ denote  $n\times n$ identity matrix and the zero element of $\R^{n}$, respectively. Whenever clear from the context, the subscript $n$ is suppressed.

\section{Problem formulation}\label{sec:Problem}
Consider an agent under observation with linear dynamics of the form \begin{equation}
	\dot{p}=q,\quad \dot{q}=Ax+Bu,\label{eq:Linear Dynamics}
\end{equation} where $ p:\R_{\geq {0}}\to\R^{n} $ denotes the generalized position, $ q:\R_{\geq {0}}\to \R^{n} $ denotes the generalized velocity, $ u:\R_{\geq {0}}\to \R^{m}$ denotes the control input, $ x\coloneqq[p^{T},q^{T}]^{T} $ denotes the state, and $ A\in \R^{n\times 2n} $ and $ B\in \R^{n\times m} $ denote the \emph{unknown} constant system matrices. Assume that the pair $ \left(A^{\prime},B^{\prime}\right) $ is controllable, where $ A^{\prime}\coloneqq\begin{bmatrix}
0_{n\times n} \quad \id_{n\times n}\\A
\end{bmatrix} $, $ B^{\prime}\coloneqq\begin{bmatrix}
0_{n\times m} \\B
\end{bmatrix} $.
The agent under observation executes a policy that minimizes the infinite horizon cost \begin{equation}
	J\left(x_{0},u\left(\cdot\right)\right)\triangleq\intop_{0}^{\infty}r\left(x\left(\tau;x_{0},u\left(\cdot\right)\right),u\left(\tau\right)\right)\textnormal{d}\tau\label{eq:cost}
\end{equation} where $ \tau\mapsto x\left(\tau;x_{0},u\left(\cdot\right)\right) $ denotes the trajectory of \eqref{eq:Linear Dynamics} under the control signal $ u\left(\cdot\right) $ starting from the initial condition $ x_{0} $ and $ r:\R^{2n}\times \R^{m}\to\R $ denotes the \emph{unknown} instantaneous cost function defined as $ r\left(x,u\right)=Q\left(x\right) + u^{T}Ru $, where $ R\in \R^{m\times m} $ is a constant positive definite matrix and $ Q:\R^{2n}\to \R^{2n} $ is a positive definite function such that an optimal policy $ u^{*}:\R^{2n}\to\R^{m} $ exists. For ease of exposition, it is further assumed that $ R=\diag\left\{r_{1},\cdots,r_{m}\right\} $, and a basis $ \sigma:\R^{2n}\to\R^{L} $ is known for $ Q $ such that for some $ W_{Q}^{*}\in \R^{L} $, \begin{equation}
	Q\left(x\right)=\left(W_{Q}^{*}\right)^{T}\sigma_{Q}\left(x\right),\forall x\in \R^{2n}.
\end{equation}

The objective of the observer is to estimate the cost function, $ r $, using measurements of the generalized position and the control input. In the following, a model-based adaptive approximate dynamic programming based approach is developed to achieve the stated objective. To facilitate model-based approximate dynamic programming, the state, $ x $, and the parameters, $ A $ and $ B $, of the UxV are estimated from the input-output measurements using a simultaneous state and parameter estimator. The state and the parameters are then utilized in an approximate dynamic programming scheme to estimate the cost.

\section{Simultaneous state and parameter estimator}\label{sec:SYSID}
The simultaneous state and parameter estimator developed by the authors in \cite{SCC.Kamalapurkar2017} is utilized in this result. This section provides a brief overview of the same for completeness. For further details, the readers are directed to \cite{SCC.Kamalapurkar2017}.

To facilitate parameter estimation, let $ A_{1},\:A_{2}\in \R^{n\times n} $ be matrices such that $ A\eqqcolon\left[A_{1},\:A_{2}\right] $. The dynamics in \eqref{eq:Linear Dynamics} can be rearranged to form the linear error system \begin{equation}
\mathcal{F}\left(t\right)=\mathcal{G}\left(t\right)\theta,\:\forall t\in\R_{\geq 0}.\label{eq:Linear Error System}
\end{equation}
In \eqref{eq:Linear Error System}, $\theta$ is a vector of unknown
parameters, defined as $\theta\triangleq\begin{bmatrix}\vecop\left(A_{1}\right)^{T} & \vecop\left(A_{2}\right)^{T} & \vecop\left(B\right)^{T}\end{bmatrix}^{T}\in\R^{2n^{2}+mn}$, where $\vecop\left(\cdot\right)$ denotes the vectorization operator and the matrices $\mathcal{F}:\R_{\geq0}\to\R^{n}$ and $\mathcal{G}:\R_{\geq0}\to\R^{n\times\left(2n^{2}+mn\right)}$
are defined as 
\begin{align*}
\mathcal{F}\left(t\right) & \triangleq\begin{cases}
\begin{gathered}p\left(t\!-\!T_{2}\!-\!T_{1}\right)\!-\!p\left(t\!-\!T_{1}\right)\\
\!+p\left(t\right)\!-\!p\left(t\!-\!T_{2}\right),
\end{gathered}
& t\!\in\!\left[T_{1}\!+\!T_{2},\infty\right),\\
0 & t<T_{1}+T_{2}.
\end{cases}\\
\mathcal{G}\left(t\right) & \triangleq\begin{bmatrix}\left(F\left(t\right)\varotimes\id_{n}\right)^{T} & \left(G\left(t\right)\varotimes\id_{n}\right)^{T} & \left(U\left(t\right)\varotimes\id_{n}\right)^{T}\end{bmatrix},
\end{align*}
where $\varotimes$ denotes the Kronecker product. The matrices $ F $, $ G $, and $ U $ are defined as $ F\left(t\right)\coloneqq \mathcal{I}p\left(t\right), \quad G\left(t\right)\coloneqq \mathcal{J}p\left(t\right)-\mathcal{J}p\left(t-T_{1}\right), \quad U\left(t\right)\coloneqq \mathcal{I}u\left(t\right)$, for $t\in\left[T_{1}+T_{2},\infty\right)$, and $F\left(t\right)=G\left(t\right)=U\left(t\right)=0$, for $ t<T_{1}+T_{2}$, where $ \mathcal{I}\coloneqq p\mapsto \intop_{t-T_{2}}^{t}\intop_{\sigma-T_{1}}^{\sigma}p\left(\tau\right)\D\tau\D\sigma $, and $ \mathcal{J}\coloneqq p\mapsto \intop_{t-T_{2}}^{t}\left(p\left(\sigma\right)\right)\D\sigma$.

For ease of exposition, it is assumed that a history stack, i.e., a set of ordered pairs $\left\{ \left(\mathcal{F}_{i},\mathcal{G}_{i}\right)\right\} _{i=1}^{M}$ such that \begin{equation} \mathcal{F}_{i}=\mathcal{G}_{i}\theta,\:\forall i\in\left\{ 1,\cdots,M\right\},\label{eq:History Stack Compatibility} \end{equation} is available a priori. A history stack $\left\{ \left(\mathcal{F}_{i},\mathcal{G}_{i}\right)\right\} _{i=1}^{M}$ is called \emph{full rank} if there exists a constant $\underline{g}\in\R$ such that \begin{equation} 0<\underline{g}<\lambda_{\min}\left\{ \mathscr{G}\right\} ,\label{eq:Rank Condition} \end{equation} where the matrix $\mathscr{G}\in\R^{\left(2n^{2}+mn\right)\times\left(2n^{2}+mn\right)}$ is defined as $\mathscr{G}\coloneqq\sum_{i=1}^{M}\mathcal{G}_{i}^{T}\mathcal{G}_{i}$. The concurrent learning update law to estimate the unknown parameters is then given by
\begin{equation}
\dot{\hat{\theta}}\left(t\right)=k_{\theta}\Gamma\left(t\right)\sum_{i=1}^{M}\mathcal{G}_{i}^{T}\left(\mathcal{F}_{i}-\mathcal{G}_{i}\hat{\theta}\left(t\right)\right),\label{eq:Theta Dynamics}
\end{equation}
where $k_{\theta}\in\R_{>0}$ is a constant adaptation gain and $\Gamma:\R_{\geq0}\to\R^{\left(2n^{2}+mn\right)\times\left(2n^{2}+mn\right)}$ is the least-squares gain updated using the update law
\begin{equation}
\dot{\Gamma}\left(t\right)=\beta_{1}\Gamma\left(t\right)-k_{\theta}\Gamma\left(t\right)\sum_{i=1}^{M}\mathcal{G}_{i}^{T}\mathcal{G}_{i}\Gamma\left(t\right).\label{eq:Gamma Dynamics}
\end{equation}
Using arguments similar to \cite[Corollary 4.3.2]{SCC.Ioannou.Sun1996}, it can be shown that provided $\lambda_{\min}\left\{ \Gamma^{-1}\left(0\right)\right\} >0$, the least squares gain matrix satisfies 
\begin{equation}
\underline{\Gamma}\id_{\left(2n^{2}+mn\right)}\leq\Gamma\left(t\right)\leq\overline{\Gamma}\id_{\left(2n^{2}+mn\right)},\label{eq:StaFGammaBound}
\end{equation}
where $\underline{\Gamma}$ and $\overline{\Gamma}$ are positive constants.

To facilitate parameter estimation based on a prediction error, a state observer is developed in the following. To facilitate the design, the dynamics in \eqref{eq:Linear Dynamics} are expressed in the form $\dot{p}\left(t\right) =q\left(t\right)$, $\dot{q}\left(t\right) =Y\left(x\left(t\right),u\left(t\right)\right)\theta$, where $Y:\R^{n}\times\R^{m}\to\R^{n\times\left(2n^{2}+mn\right)}$ is defined as 
\[
Y\left(x,u\right)=\begin{bmatrix}\left(p\varotimes\id_{n}\right)^{T} & \left(q\varotimes\id_{n}\right)^{T} & \left(u\varotimes\id_{n}\right)^{T}\end{bmatrix}.
\]
The adaptive state observer is then designed as
\begin{align}
\dot{\hat{p}}\left(t\right) & =\hat{q}\left(t\right),\quad\hat{p}\left(0\right)=p\left(0\right),\nonumber \\
\dot{\hat{q}}\left(t\right) & =Y\left(x\left(t\right),u\left(t\right)\right)\hat{\theta}\left(t\right)+\nu\left(t\right),\quad\hat{q}\left(0\right)=0,\label{eq:State observer}
\end{align}
where $\hat{p}:\R_{\geq 0}\to\R^{n}$, $\hat{q}:\R_{\geq 0}\to\R^{n}$, $\hat{x}:\R_{\geq 0}\to\R^{n}$, and $\hat{\theta}:\R_{\geq 0}\to\R^{n}$ are estimates of $p$, $q$, $x$, and $\theta$, respectively, $\nu$ is the feedback component of the identifier, to be designed later, and the prediction error $\tilde{p}:\R_{\geq 0}\to\R^{n}$ is defined as $\tilde{p}\left(t\right)=p\left(t\right)-\hat{p}\left(t\right).$

The update law for the generalized velocity estimate depends on the entire state $x$. However, using the structure of the matrix $Y$ and integrating by parts, the observer can be implemented without using generalized velocity measurements. Using an integral form of (\ref{eq:State observer}), the update law in (\ref{eq:State observer}) can be implemented without generalized velocity measurements as
\begin{multline}
\hat{q}\left(t\right)=\intop_{0}^{t}\left(u\left(\tau\right)\varotimes\id_{n}\right)^{T}\vecop\left(\hat{B}\left(\tau\right)\right)\D\tau+\intop_{0}^{t}\nu\left(\tau\right)\D\tau\\
+\hat{q}\left(0\right)+\intop_{0}^{t}\!\!\left(p\left(\tau\right)\varotimes\id_{n}\right)^{T}\!\!\left(\!\vecop\left(\hat{A}_{1}\left(\tau\right)\right)\!-\!\vecop\left(\dot{\hat{A}}_{2}\left(\tau\right)\right)\!\right)\D\tau\\
+\!\left(p\left(t\right)\!\varotimes\!\id_{n}\right)^{T}\!\!\vecop\!\left(\!\!\hat{A}_{2}\left(t\right)\!\right)\!-\!\left(p\left(0\right)\!\varotimes\!\id_{n}\right)^{T}\!\!\vecop\!\left(\!\!\hat{A}_{2}\left(0\right)\!\right)\label{eq:IntegralUpdate}
\end{multline}

To facilitate the design of the feedback component $\nu$, let 
\begin{equation}
r\left(t\right)=\tilde{q}\left(t\right)+\alpha\tilde{p}\left(t\right)+\eta\left(t\right),\label{eq:r}
\end{equation}
where $\alpha>0$ is a constant observer gain and the signal $\eta$ is added to compensate for the fact that the generalized velocity state, $q$, is not measurable. Based on the subsequent stability analysis, the signal $\eta$ is designed as the output of the dynamic filter 
\begin{align}
\dot{\eta}\left(t\right) & =-\beta\eta\left(t\right)-kr\left(t\right)-\alpha\tilde{q}\left(t\right),\quad\eta\left(0\right)=0,\label{eq:eta Update}
\end{align}
and the feedback component $\nu$ is designed as 
\begin{equation}
\nu\left(t\right)=\tilde{p}\left(t\right)-\left(k+\alpha+\beta\right)\eta\left(t\right),\label{eq:Feedback}
\end{equation}
where $\beta>0$ and $k>0$ are constant observer gains. The design of the signals $\eta$ and $\nu$ to estimate the state from output measurements is inspired by the $p-$filter (cf. \cite{SCC.Xian.Queiroz.ea2004}). Similar to the update law for the generalized velocity, using the the fact that $\tilde{p}\left(0\right)=0$, the signal $\eta$ can be implemented using the integral form
\begin{equation}
\eta\left(t\right)=-\intop_{0}^{t}\left(\beta+k\right)\eta\left(\tau\right)\D\tau-\intop_{0}^{t}k\alpha\tilde{p}\left(\tau\right)\D\tau-\left(k+\alpha\right)\tilde{p}\left(t\right).\label{eq:IntegralUpdateEta}
\end{equation}

Using a Lyapunov-based analysis, it can be shown that the developed parameter and state estimation results in exponential convergence of the state and parameter estimation errors to zero. For a detailed analysis of the developed state and parameter estimator, see \cite{SCC.Kamalapurkar2017}.
\section{Inverse Bellman Error}\label{sec:IBE}
Since the agent under observation makes optimal decisions, and since the Hamiltonian $H:\R^{2n}\times\R^{2n}\times \R^{m}\to\R$, defined as $H\left(x,y,u\right)\triangleq y^{T}\left(A^{\prime}x+B^{\prime}u\right)+r\left(x,u\right)$, is convex in $ u $, the control signal, $ u\left(\cdot\right) $, and the state, $ x\left(\cdot\right) $, satisfy the Hamilton-Jacobi-Bellman equation \begin{equation}
	H\left(x\left(t\right),\nabla_{x}\left(V^{*}\left(x\left(t\right)\right)\right)^{T},u\left(t\right)\right)=0,\forall t\in \R_{\geq 0},\label{eq:inverse HJB}
\end{equation}
where $ V^{*}:\R^{2n}\to \R $ denotes the unknown optimal value function. The objective of inverse reinforcement learning is to generate an estimate of the unknown cost function, $ r $. To facilitate estimation of the cost function, let $ \hat{V}:\R^{2n}\times \R^{P} \to \R$, $ \left(x,\hat{W}_{V}\right)\mapsto \hat{W}_{V}^{T}\sigma_{V}\left(x\right) $ be a parametric estimate of the optimal value function, where $ \hat{W}_{V}\in \R^{P} $ are unknown parameters, and $ \sigma_{V}:\R^{2n}\to\R^{P} $ are known continuously differentiable features. Assume that given any compact set $ \chi\subset\R^{2n} $ and a constant $ \overline{\epsilon}>0 $, sufficiently many features can be selected to ensure the existence of ideal parameters $ W_{V}^{*}\in \R^{P} $ such that the error $ \epsilon:\R^{2n}\to\R $, defined as $ \epsilon\left(x\right) \coloneqq V\left(x\right)-\hat{V}\left(x,W_{V}^{*}\right)$, satisfies $ \sup_{x\in \chi}\left|\epsilon\left(x\right)\right|<\overline{\epsilon} $ and $ \sup_{x\in \chi}\left|\nabla_{x}\epsilon\left(x\right)\right|<\overline{\epsilon} $. Using the estimates $ \hat{A}_{1} $, $ \hat{A}_{2} $, $ \hat{B} $, $ \hat{W}_{V} $, $ \hat{W}_{Q} $, and $ \hat{W}_{R} $ of the parameters $ A_{1} $, $ A_{2} $, $ B $, $ W_{V}^{*} $, $ W_{Q}^{*} $, and $ W_{R}\coloneqq\left[r_{1},\cdots,r_{m}\right]^{T} $, respectively, and the estimate $ \hat{x} $ of the state, $ x $, in \eqref{eq:inverse HJB}, the inverse Bellman error $ \delta^{\prime}:\R^{2n}\times \R^{m}\times\R^{L+P+m}\times \R^{2n^{2}+mn}\to\R $ is obtained as \begin{align}
	\delta^{\prime}\left(\hat{x},u,\hat{W},\hat{\theta}\right)=&\hat{W}_{V}^{T}\nabla_{x}\sigma_{V}\left(\hat{x}\right)\left(\hat{A}^{\prime}\hat{x}+\hat{B}^{\prime}u\right)+\hat{W}_{Q}^{T}\sigma_{Q}\left(\hat{x}\right)\nonumber\\
	&+\hat{W}_{R}^{T}\sigma_{u}\left(u\right),
\end{align}where $ \sigma_{u}\left(u\right)\coloneqq\left[u_{1}^{2},\cdots,u_{m}^{2}\right]$, $ \hat{A}^{\prime} \coloneqq\begin{bmatrix}
0_{n\times n} & \id_{n\times n}\\
\hat{A}_{1} & \hat{A_{2}}
\end{bmatrix}$, and $ \hat{B}^{\prime}\coloneqq\begin{bmatrix}
0_{n\times m}\\\hat{B}
\end{bmatrix} $. Rearranging, \begin{equation}
\delta^{\prime}\left(\hat{x},u,\hat{W}^{\prime},\hat{\theta}\right)=\left(\hat{W}^{\prime}\right)^{T}\sigma^{\prime}\left(\hat{x},u,\hat{\theta}\right),\label{eq:inverse BE}
\end{equation}where $ \hat{W}^{\prime} \coloneqq \left[\hat{W}_{V};\hat{W}_{Q};\hat{W}_{R}\right] $, $ \sigma^{\prime}\left(\hat{x},u,\hat{\theta}\right)\coloneqq\left[\nabla_{x}\sigma_{V}\left(\hat{x}\right)\left(\hat{A}^{\prime}\hat{x}+\hat{B}^{\prime}u\right);\sigma_{Q}\left(\hat{x}\right);\sigma_{u}\left(u\right)\right] $. The following section details the developed model-based inverse reinforcement learning algorithm.

\section{Inverse Reinforcement Learning}\label{sec:IRL}
The IRL problem can be solved by computing the estimates $\hat{W}$ that minimize the inverse Bellman error in \eqref{eq:inverse BE}. To facilitate the computation, the values of $ \hat{x} $, $ u $, and $ \hat{\theta} $ are recorded at time instances $ \left\{t_{i}<t\right\}_{i=1}^{N} $ to generate the values $\left\{\hat{\sigma}_{t}^{\prime}\left(t_{i}\right)\right\}_{i=1}^{N}$, where $ N\in\n $, $ N>>L+P+m $, and  $ \hat{\sigma}_{t}^{\prime}\left(t\right)\coloneqq\sigma^{\prime}\left(\hat{x}\left(t\right),u\left(t\right),\hat{\theta}\left(t\right)\right) $. The data in the history stack can be collected in a matrix form to yield \begin{equation}
	\Delta^{\prime} = \hat{\Sigma}^{\prime} \hat{W}^{\prime},\label{homogeneous}
\end{equation}where $ \Delta^{\prime}\coloneqq\left[\delta^{\prime}_{t}\left(t_{1}\right);\cdots;\delta^{\prime}_{t}\left(t_{N}\right)\right] $, $ \delta^{\prime}_{t}\left(t\right)\coloneqq \delta^{\prime}\left(\hat{x}\left(t\right),u\left(t\right),\hat{W}^{\prime},\hat{\theta}\left(t\right)\right)$, and $ \hat{\Sigma}^{\prime}\coloneqq\left[\left(\hat{\sigma}_{t}^{\prime}\right)^{T}\left(t_{1}\right);\cdots;\left(\hat{\sigma}_{t}^{\prime}\right)^{T}\left(t_{N}\right)\right] $. Note that the solution $\hat{W}^{\prime}=0$ trivially minimizes $\Delta^{\prime}$, which is to say that if the cost function is identically zero then every policy is optimal. Hence, as stated, the IRL problem is clearly ill-posed. In fact, the cost functions $r\left(x,u\right)$ and $Kr\left(x,u\right)$, where $K$ is a positive constant, result in identical optimal policies and state trajectories. Hence, even if the trivial solution is discarded, the cost function can only be identified up to multiplication by a positive constant using the trajectories $x\left(\cdot\right)$ and $u\left(\cdot\right)$.

To remove the aforementioned ambiguity without loss of generality, the first element of $ \hat{W}_{R} $ is assumed to be known. The inverse BE in \eqref{eq:inverse BE} can then be expressed as \begin{equation}
	\delta^{\prime}\left(\hat{x},u,\hat{W},\hat{\theta}\right)=\hat{W}^{T}\sigma^{\prime\prime}\left(\hat{x},u,\hat{\theta}\right) + r_{1}\sigma_{u1}\left(u\right),
\end{equation}where $ \sigma_{ui}\left(u\right) $ denotes the $ i $\textsubscript{th} element of the vector $ \sigma_{u}\left(u\right) $, the vector $ \sigma_{u}^{-} $ denotes $ \sigma_{u} $, with the first element removed, and $ \sigma^{\prime\prime}\left(\hat{x},u,\hat{\theta}\right)\coloneqq\left[\nabla_{x}\sigma_{V}\left(\hat{x}\right)\left(\hat{A}^{\prime}\hat{x}+\hat{B}^{\prime}u\right);\sigma_{Q}\left(\hat{x}\right);\sigma_{u}^{-}\left(u\right)\right] $.

The closed-form optimal controller corresponding to \eqref{eq:cost} provides the relationship\begin{equation}
	-2Ru\left(t\right)=\left(B^{\prime}\right)^{T}\nabla_{x}\sigma_{V}\left(x\left(t\right)\right)W_{V}^{*}+\left(B^{\prime}\right)^{T}\nabla_{x}\epsilon\left(x\left(t\right)\right),
\end{equation}which can be expressed as \begin{align*}
	-2r_{1}u_{1}\left(t\right) + \Delta_{u1}&=\sigma_{B1}\hat{W}_{V}\\
	\Delta_{u^{-}}&=\sigma_{B}^{-}\hat{W}_{V}+2\diag\left(u_{2},\cdots,u_{m}\right)\hat{W}_{R}^{-},
\end{align*}where $ \sigma_{B1} $ and $ u_{1} $ denote the first rows and $ \sigma_{B}^{-}  $ and $ u^{-} $ denote all but the first rows of $\sigma_{B}\coloneqq \left(B^{\prime}\right)^{T}\nabla_{x}\sigma_{V}\left(x\right) $ and $ u $, respectively, and $ R^{-}\coloneqq\diag\left(\left[r_{2},\cdots,r_{m}\right]\right) $. For notational brevity let $ \sigma\coloneqq\left[\sigma^{\prime\prime},\, \begin{bmatrix}
\sigma_{B}^{T}\\\Theta
\end{bmatrix}\right]$, where \[ \Theta\coloneqq\left[0_{m\times 2n}, \,\,\begin{bmatrix}
0_{1\times m-1}\\2\diag\left(\left[u_{2},\cdots,u_{m}\right]\right)
\end{bmatrix}\right]^{T} \]

The history stack can then be utilized to generate the linear system \begin{equation}
	-\Sigma_{u1}=\hat{\Sigma}\hat{W}-\Delta^{\prime},
\end{equation}where $ \hat{W} \coloneqq \left[\hat{W}_{V};\hat{W}_{Q};\hat{W}_{R}^{-}\right] $, $ \hat{\Sigma}\coloneqq\left[\hat{\sigma}_{t}^{T}\left(t_{1}\right);\cdots;\hat{\sigma}_{t}^{T}\left(t_{N}\right)\right] $, and $ \Sigma_{u1}\coloneqq\left[\sigma_{u1}^{\prime}\left(u\left(t_{1}\right)\right);\cdots;\sigma_{u1}^{\prime}\left(u\left(t_{N}\right)\right)\right] $, where $ \hat{\sigma}_{t}\left(\tau\right)\coloneqq\sigma\left(\hat{x}\left(\tau\right),u\left(\tau\right),\hat{\theta}\left(\tau\right)\right) $, $ \sigma_{u1}^{\prime}\coloneqq \left[\sigma_{u1};2r_{1}u_{1};0_{\left(m-1\right)\times 1}\right]$, and the vector $ \hat{W}_{R}^{-} $ denotes $ \hat{W}_{R} $ with the first element removed.

At any time instant $ t $, provided the history stack $ \mathcal{G}\left(t\right) $ satisfies \begin{equation}
	\textnormal{rank}\left(\hat{\Sigma}\right)=L+P+m-1,\label{eq:Rank Condition}
\end{equation} then a least-squares estimate of the weights can be obtained as\begin{equation}
	\hat{W}\left(t\right)=-\left(\hat{\Sigma}^{T}\hat{\Sigma}\right)^{-1}\hat{\Sigma}^{T}\Sigma_{u1}.\label{eq:Least Squares}
\end{equation} To improve numerical stability of the least-squares solution, the data recoded in the history stack is selected to maximize the condition number of $ \hat{\Sigma} $ while ensuring that the vector $ \Sigma_{u1} $ remains nonzero. The data selection algorithm is detailed in Fig. \ref{alg:DataSelect}.
\begin{figure}
	\begin{algorithmic}[1]		
		\If{an observed, estimated or queried data point $ \left(x^{*},u^{*}\right) $ is available at $ t=t^{*} $}		
		\If{the history stack is not full}		
		\State add the data point to the history stack
		\ElsIf{$ \kappa\!\left(\!\!\left(\hat{\Sigma}\left(i\leftarrow*\right)\right)^{T}\!\!\left(\hat{\Sigma}\left(i\leftarrow*\right)\right)\!\!\right)\!<\!\xi_{1}\kappa\left(\hat{\Sigma}^{T}\hat{\Sigma}\right)$,\break\hspace*{1.2em} for some $ i $, and $\left\Vert \Sigma_{u1}\left(i\leftarrow*\right) \right\Vert\geq\xi_{2}$}		
		\State add the data point to the history stack
		\State $ \varpi \leftarrow 1 $
		\Else		
		\State discard the data point
		\State $ \varpi \leftarrow 0 $
		\EndIf
		\EndIf
	\end{algorithmic}
	
	\caption{\label{alg:DataSelect}Algorithm for selecting data for the history stack. The constants $\xi_{1}\geq0$ and $ \xi_{2}>0 $ are tunable thresholds. The operator $ \kappa\left(\cdot\right) $ denotes the condition number of a matrix. For the matrix $ \hat{\Sigma} = \left[\hat{\sigma}_{t}^{T}\left(t_{1}\right);\cdots;\hat{\sigma}_{t}^{T}\left(t_{i}\right);\cdots;\hat{\sigma}_{t}^{T}\left(t_{N}\right)\right] $, $ \Sigma\left(i\leftarrow*\right) \coloneqq \left[\hat{\sigma}_{t}^{T}\left(t_{1}\right);\cdots;\hat{\sigma}_{t}^{T}\left(t^{*}\right);\cdots;\hat{\sigma}_{t}^{T}\left(t_{N}\right)\right]$ and for the vector $ \Sigma_{u1} = \left[\sigma_{u1}\left(u\left(t_{1}\right)\right);\cdots;\sigma_{u1}\left(u\left(t_{i}\right)\right);\cdots;\sigma_{u1}\left(u\left(t_{N}\right)\right)\right] $, $ \Sigma_{u1}\left(i\leftarrow*\right) \coloneqq \left[\sigma_{u1}\left(u\left(t_{1}\right)\right);\cdots;\sigma_{u1}\left(u\left(t^{*}\right)\right);\cdots;\sigma_{u1}\left(u\left(t_{N}\right)\right)\right]$.}
\end{figure}
\section{Purging to Exploit Improved State and Parameter Estimates}\label{sec:Purge}
Since the matrix $ \Sigma $ is a function of the state and parameter estimates, the accuracy of the least-squares solution in \eqref{eq:Least Squares} depends on the accuracy of the state and parameter estimates recoded in $ \mathcal{G} $. The state and parameter estimates are likely to be poor during the transient phase of the estimator dynamics. As a result, a least-squares solution computed using data recorded during the transient phase of the estimator may be inaccurate. Based on the observation that the state and the parameter estimates exponentially decay to the origin, a purging algorithm is developed in the following to remove erroneous state and parameter estimates from the history stack.

Numerical differentiation is utilized to to gauge the quality of the state estimates. Given a constant $ T>0 $, and the position measurements over the interval $ \left[0,t\right] $, noncausal numerical smoothing methods can be used to generate an accurate estimate $ \dot{\overline{p}}\left(t-T\right) $ of the velocity. The signal $ \eta_{1}\left(t\right)\coloneqq \overline{x}^{T}S_{1}\overline{x}$ is used as an indicator of the quality of the state estimates, where $ \overline{x}\coloneqq\left[\tilde{p}\left(t\right);\dot{\overline{p}}\left(t-T\right)\right] $, and $ S_{1}\in \R^{2n\times2n} $ is a positive semidefinite matrix.

To gauge the quality of the parameter estimates, at each time instance $ t $, the model $ \dot{p}\left(\tau\right)=q\left(\tau\right),\quad \dot{q}\left(\tau\right)=\hat{A}x\left(\tau\right)+\hat{B}u\left(\tau\right), $ is simulated over $ \tau\in\left[t-T,t\right] $, using the initial condition $ x\left(t-T\right) = \left[p\left(t-T\right);\dot{\overline{p}}\left(t-T\right)\right]$, to generate the trajectory $ \overline{p}:\left[t-T,t\right]\to\R^{n} $. The signal $ \eta_{2}\left(t\right)\coloneqq\intop_{t-T}^{t}\left(\overline{p}\left(\tau\right)\right)^{T}S_{2}\overline{p}\left(\tau\right)\D\tau$ is used as an indicator of the quality of the parameter estimates, where $ S_{2}\in \R^{n\times n} $ is a positive semidefinite matrix. The composite signal $ \eta\left(t\right)\coloneqq\eta_{1}\left(t\right)+\eta_{2}\left(t\right) $ is used as an indicator of the quality of the simultaneous state and parameter estimator. 

The indicator developed above is used to purge and update the history stack and to update the estimate $ \hat{W} $ according to the algorithm detailed in Fig. \ref{alg:Purging}. The algorithm begins with an empty history stack and an initial estimate of the weights $ W_{0} $. Values of $ \hat{x} $, $ u $, $ \hat{\theta} $, and $ \eta $ are recorded in the history stack using the algorithm detailed in Fig. \ref{alg:DataSelect}, where $ \eta\left(t\right) $ is assumed to be infinite for $ t<T $. The estimate $ \hat{W} $ is held at the initial guess until the history stack is full. Then, it is updated using $ \eqref{eq:Least Squares} $ every time a new data point is added to the history stack.

Communication with the entity under observation, wherever possible can be easily incorporated in the developed framework. In the query-based implementation of the developed algorithm, the observed input-output trajectories are utilized to learn the dynamics of the UxV. Instead of using the estimated state and control trajectories for cost estimation, control actions, $ u_{i} $ of the entity under observation in response to randomly selected states, $ x_{i} $, are queried. If the queried state-input pair improves the condition number of the history stack then it is stored in the history stack and utilized for cost estimation.
\begin{figure}
	\begin{algorithmic}[1]
		\State 	$ \hat{W}\left(0\right)\leftarrow W_{0} $, $ s\leftarrow 0 $
		\If{$\kappa\left(\hat{\Sigma}^{T}\hat{\Sigma}\right)<\underline{\kappa_{1}}$ and $ \varpi=1 $}		
		\State $ \hat{W}\left(t\right) \leftarrow -\left(\hat{\Sigma}^{T}\hat{\Sigma}\right)^{-1}\hat{\Sigma}^{T}\Sigma_{u1}$
		\Else
		\State Hold $ \hat{W} $ at the previous value
		\EndIf
		\If{$ \kappa\left(\hat{\Sigma}^{T}\hat{\Sigma}\right)<\underline{\kappa_{2}}$ and $ \eta\left(t\right)<\overline{\eta}\left(t\right) $}		
		\State empty the history stack
		\State $ s\leftarrow s+1 $
		\EndIf
	\end{algorithmic}
	
	\caption{\label{alg:Purging}Algorithm for updating the weights and the history stack. The constants $\underline{\kappa_{1}}>0$ and $ \underline{\kappa_{2}}>0 $ are tunable thresholds, the index $ s $ denotes the number of times the history stack was purged, and $ \overline{\eta}\left(t\right)\coloneqq\min\left\{\eta\left(t_{1}\right),\cdots,\eta\left(t_{M}\right)\right\} $.}
\end{figure}

\section{Analysis}\label{sec:Ana}

A detailed analysis of the simultaneous state and parameter estimator is excluded for brevity, and is available in \cite{SCC.Kamalapurkar2017}. To facilitate the analysis of the IRL algorithm, let $ \Sigma\coloneqq\left[\sigma\left(x\left(t_{1}\right),u\left(t_{1}\right),\theta\right);\cdots;\sigma\left(x\left(t_{M}\right),u\left(t_{M}\right),\theta\right)\right] $ and let $ \hat{W}^{*} $ denote the least-squares solution of $ \Sigma\hat{W}=-\Sigma_{u1} $. Furthermore, let $ W $ denote an appropriately scaled version of the ideal weights, i.e, $W\coloneqq \nicefrac{W}{r_{1}}$. Provided the rank condition in \eqref{eq:Rank Condition} is satisfied, the inverse HJB equation in \ref{eq:inverse HJB} implies that $ \Sigma W=-\Sigma_{u1}-E $, where $ E\coloneqq[\nabla_{x}\epsilon\left(x\left(t_{1}\right)\right)\left(Ax\left(t_{1}\right)+Bu\left(t_{1}\right)\right)$; $\cdots$; $\nabla_{x}\epsilon\left(x\left(t_{M}\right)\right)\left(Ax\left(t_{M}\right)+Bu\left(t_{M}\right)\right)] $. That is, $ \left\Vert  W+\left(\Sigma^{T}\Sigma\right)^{-1}\Sigma^{T}\Sigma_{u1} \right\Vert\leq\left\Vert \left(\Sigma^{T}\Sigma\right)^{-1}\Sigma^{T}E \right\Vert $. Since $ \hat{W}^{*} $ is a least squares solution, $ \left\Vert W-\hat{W}^{*} \right\Vert\leq\left\Vert \left(\Sigma^{T}\Sigma\right)^{-1}\Sigma^{T}E \right\Vert $.

Let $ \hat{\Sigma}_{s} $, $ \Sigma_{u1_{s}} $, and $ \hat{W}_{s} $ denote the regression matrices and the weight estimates corresponding to the s\textsuperscript{th} history stack, respectively, and let $ \Sigma_{s} $ denote the ideal regression matrix where $ \hat{x}\left(t_{i}\right) $ and $ \hat{\theta}\left(t_{i}\right) $ in $ \hat{\Sigma}_{s} $ are replaced with the corresponding ideal values $ x\left(t_{i}\right) $ and $ \theta $. Let $ \hat{W}^{*}_{s} $ denote the least-squares solution of $ \Sigma_{s}\hat{W}=-\Sigma_{u1_{s}} $. Provided  $ \hat{\Sigma}_{s} $ satisfies the rank condition in \eqref{eq:Rank Condition}, then $ \left\Vert W-\hat{W}^{*}_{s} \right\Vert\leq\left\Vert \left(\Sigma^{T}_{s}\Sigma_{s}\right)^{-1}\Sigma_{s}^{T}E \right\Vert $. Furthermore, $ \hat{W}_{s}-\hat{W}^{*}_{s} = \left(\left(\left(\hat{\Sigma}^{T}_{s}\hat{\Sigma}_{s}\right)^{-1}\hat{\Sigma}_{s}^{T}\right)-\left(\left(\Sigma^{T}_{s}\Sigma_{s}\right)^{-1}\Sigma_{s}^{T}\right)\right)\Sigma_{u1_{s}}$ Since the estimates $ \hat{x} $ and $ \hat{\theta} $ exponentially converge to $ x $ and $ \theta $, respectively, the function $ \left(x,\theta\right)\mapsto \sigma\left(x,u,\theta\right) $ is continuous for all $ u $, and under the rank condition in \eqref{eq:Rank Condition}, the function $ \Sigma\mapsto\left(\Sigma^{T}\Sigma\right)^{-1}\Sigma^{T} $ is continuous, it can be concluded that $ \hat{W}_{s}\to\hat{W}^{*}_{s} $ as $ s\to\infty $, and hence, the error between the estimates $ \hat{W}_{s} $ and the ideal weights $ W $ is $ O\left(\overline{\epsilon}\right) $ as $ s\to\infty $.

\section{Simulation}
To verify the performance of the developed method, a linear quadratic optimal control problem is selected where \begin{align*}
	A&=\begin{bmatrix}
	1 &1 &-1 &1\\5&1&1&1
	\end{bmatrix},\\B&=\begin{bmatrix}
	1&3\\0&1
	\end{bmatrix}.
\end{align*}
The weighing matrices in the cost function are selected as $ Q=\diag\left(\left[1,\,2,\,3,\,6\right]\right) $ and $ R=\left[20,\,10\right] $, where $ R\left(1,1\right) $ is assumed to be known. The observed input-output trajectories, along with a prerecorded history stack are used to implement the simultaneous state and parameter estimation algorithm in Section \ref{sec:SYSID}. The design parameters in the system identification algorithm are selected using trial and error as $ M=150 $, $ T_{1}=1s $, $ T_{2}=0.8s $$ k=100 $, $ \alpha=20 $, $ \beta=10 $, $ \beta_{1}=5 $,  $ k_{\theta}=0.3/M $, and $ \Gamma\left(0\right)=0.1*\id_{L+P+m-1} $.

\begin{figure}
	\centering
	\includegraphics[width=1\columnwidth]{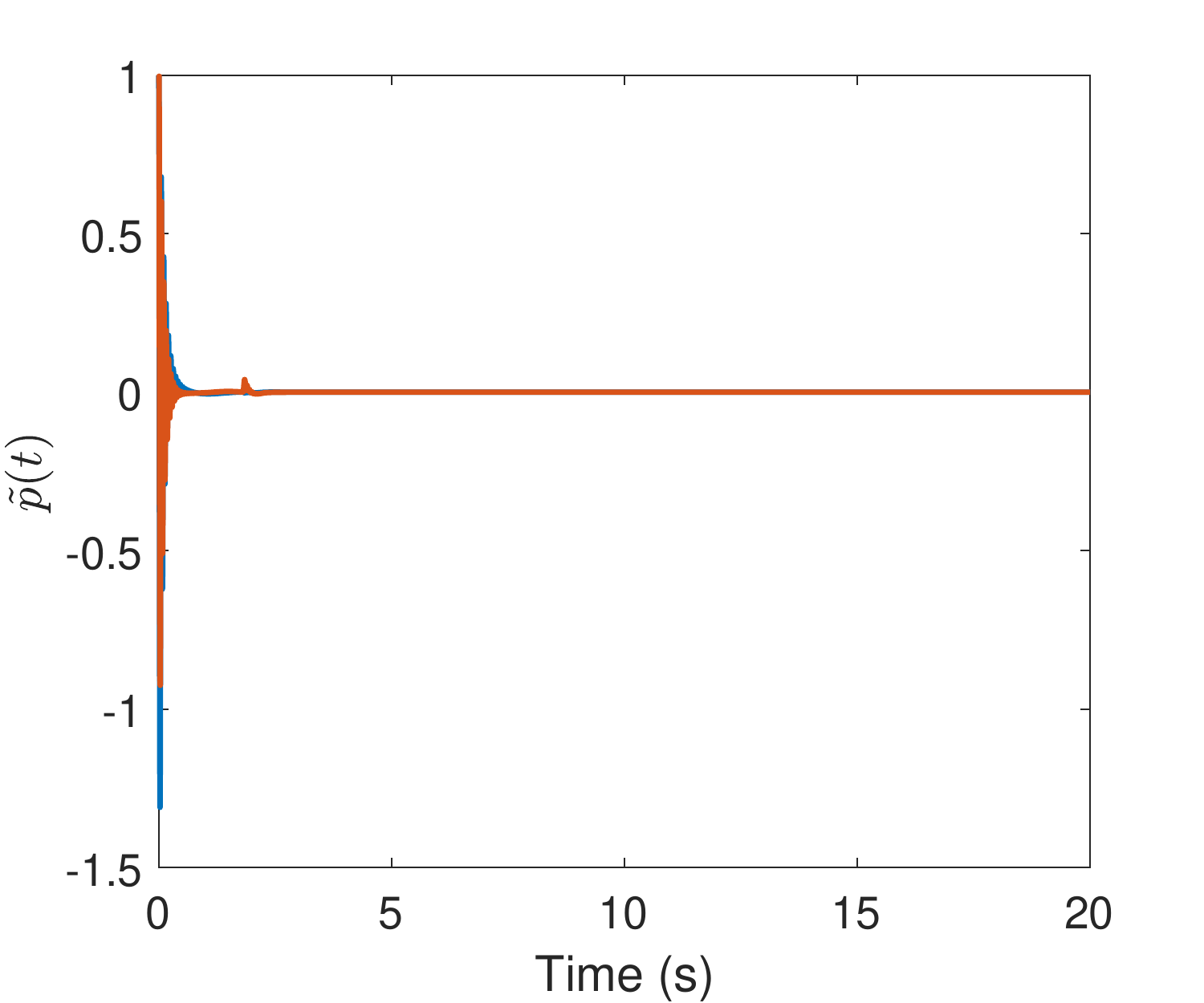}
	\caption{\label{fig:ptilde}Generalized position estimation error.}
\end{figure}\begin{figure}
\centering
\includegraphics[width=1\columnwidth]{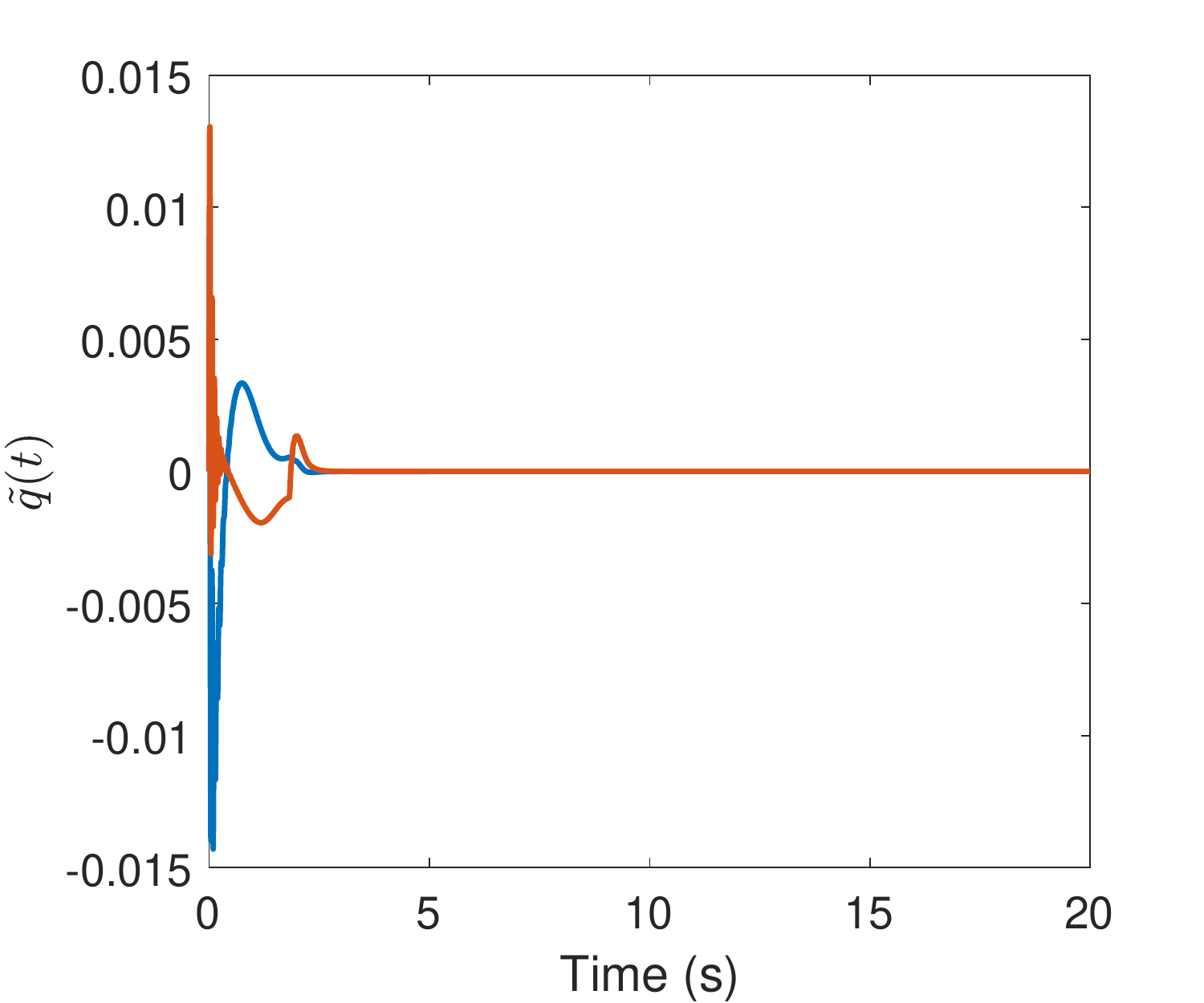}
\caption{\label{fig:qtilde}Generalized velocity estimation error.}
\end{figure}\begin{figure}
\centering
\includegraphics[width=1\columnwidth]{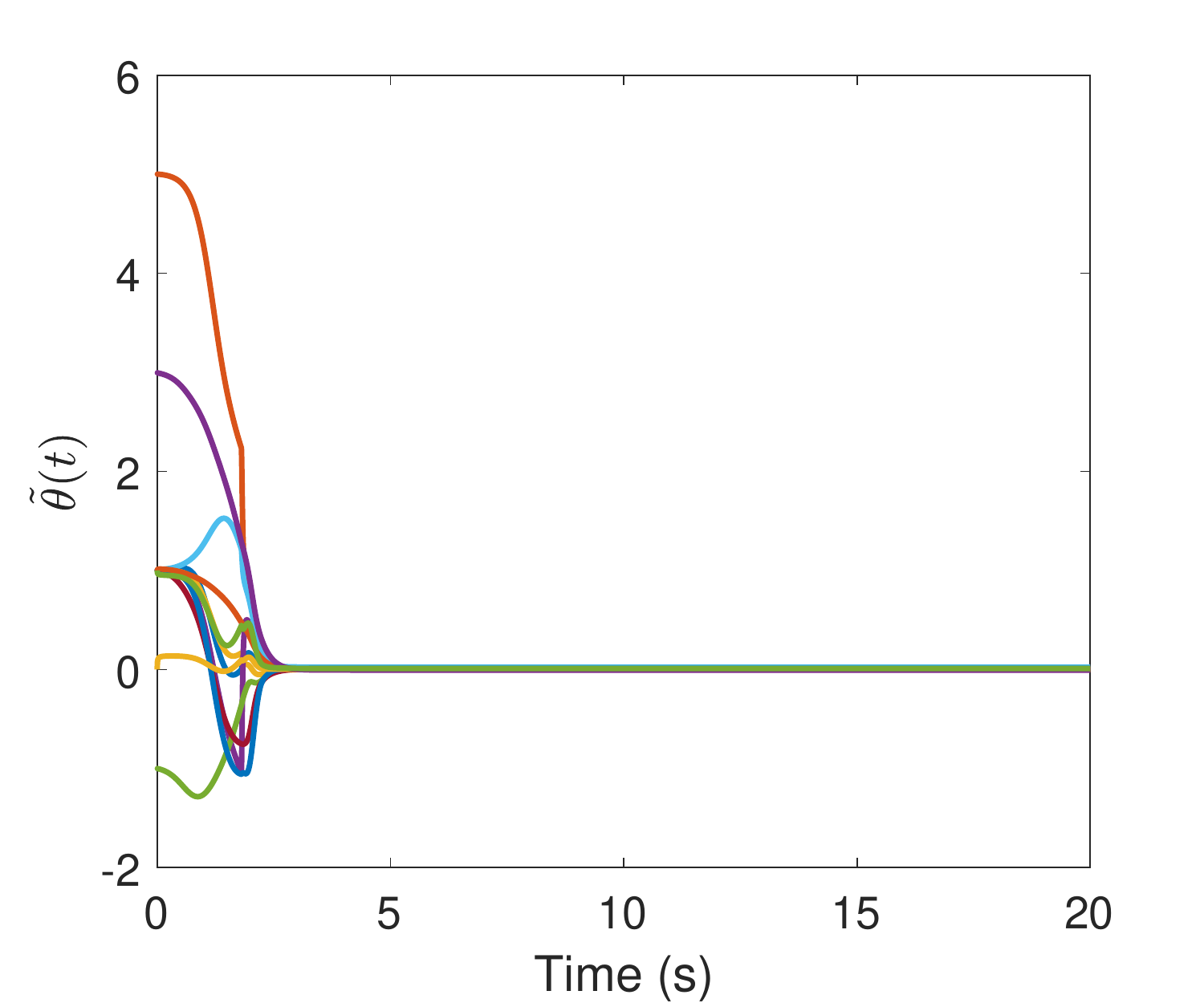}
\caption{\label{fig:thetatilde}Estimation error for the unknown parameters in the system dynamics.}
\end{figure}\begin{figure}
\centering
\includegraphics[width=1\columnwidth]{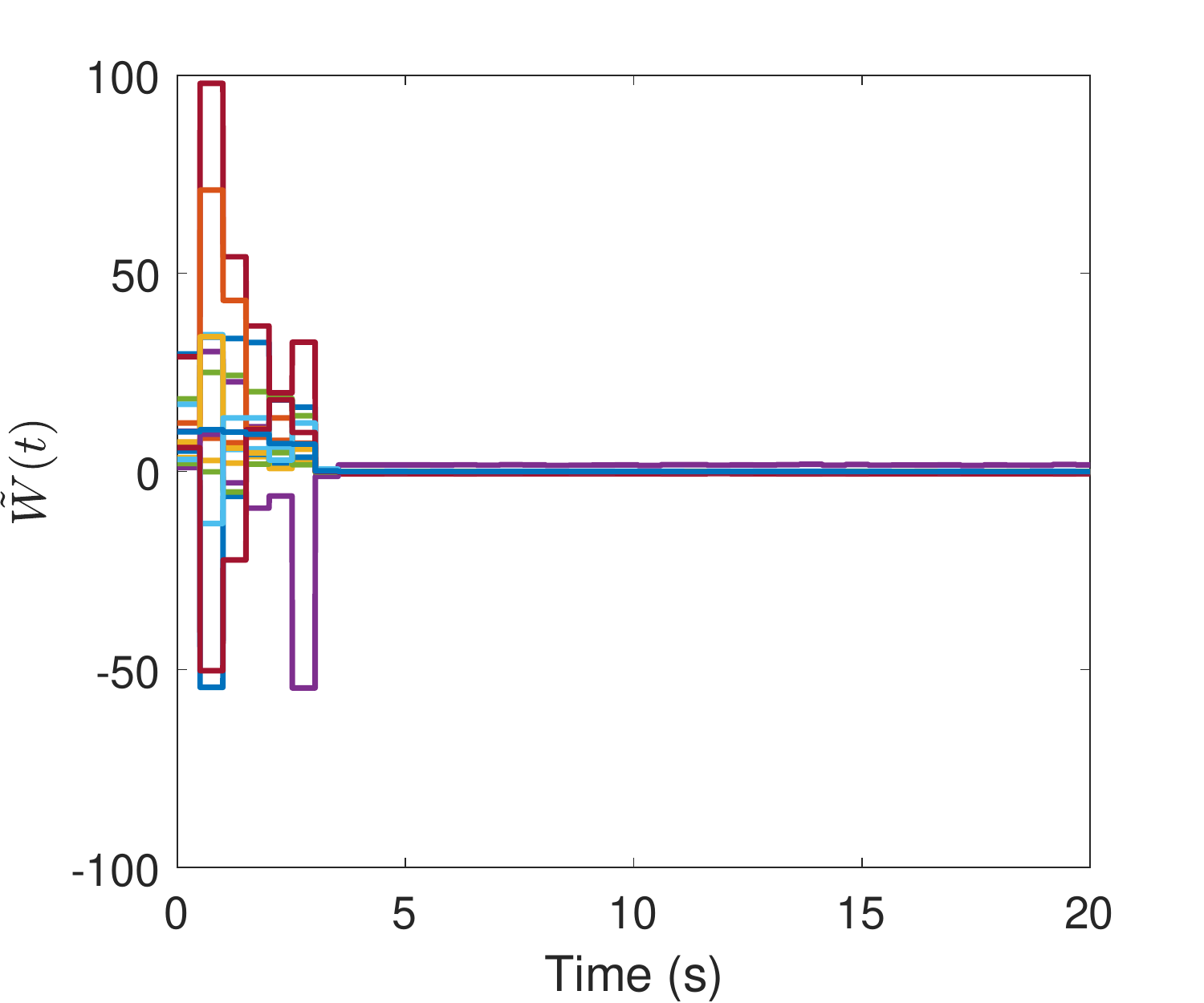}
\caption{\label{fig:WTilde}Estimation error for the unknown parameters in the cost function.}
\end{figure}

The behavior of the system under the optimal controller $ u\left(t\right)=R^{-1}\left(B^{\prime }\right)^{T}Px\left(t\right)$ is observed, where $ P\in \R^{2n\times2n} $ is the solution to the algebraic Riccati equation corresponding to \eqref{eq:cost}. At each time step, a random state vector $ x^{*} $ is selected and the optimal action $ u^{*} $ corresponding to the random state vector is queried from the entity under observation. The queried state-action pairs $ \left(x^{*},u^{*}\right) $ are utilized in conjunction with the estimated state-action pairs $ \left(\hat{x\left(t\right)},u\left(t\right)\right) $ to implement the IRL algorithm developed in Section \eqref{sec:IRL}.

Figs. \ref{fig:ptilde} and \ref{fig:qtilde} demonstrate the performance of the developed state estimator and Fig. \ref{fig:thetatilde} illustrates the performance of the developed parameter estimator. The estimation errors in the generalized position, the generalized velocity, and the unknown plant parameters exponentially decay to the origin. Fig. \ref{fig:WTilde} indicates that the developed IRL technique can be successfully utilized to estimate the cost function of an entity under observation.

\section{Conclusion}\label{sec:Con}
A data-driven inverse reinforcement learning technique is developed for a class of linear systems to estimate the cost function of an agent online, using input-output measurements. A simultaneous state and parameter estimator is utilized to facilitate output-feedback inverse reinforcement learning, and cost function estimation is achieved up to multiplication by a constant. A purging algorithm is utilized to update the stored state and parameter estimates and bounds on the cost estimation error are obtained.
\bibliographystyle{ieeetran}
\bibliography{scc,sccmaster}
\end{document}

%% file: Macros.tex
\global\long\def\D{\,\mathrm{d}}

\global\long\def\R{\mathbb{R}}

\global\long\def\B{\operatorname{B}}

\global\long\def\n{\mathbb{N}}

\global\long\def\diag{\operatorname{diag}}

\newcommandx\fid[5][usedefault, addprefix=\global, 1=, 2=, 3=, 4=, 5=]{\tensor*[_{#3}^{#2}]{#1}{_{#5}^{#4}}}

\global\long\def\vecop{\operatorname{vec}}

\global\long\def\id{\operatorname{I}}

%% file: Theorems.tex
\providecommand{\theoremname}{Theorem}

\newtheorem*{thm*}{\protect\theoremname}

\providecommand{\conjucturename}{Conjucture}

\newtheorem*{conjucture*}{\protect\conjucturename}

\providecommand{\questionname}{Question}
\theoremstyle{definition}

\theoremstyle{definition}
\newtheorem*{question*}{\protect\questionname}

\providecommand{\lemmaname}{Lemma}

\newtheorem*{lem*}{\protect\lemmaname}

\providecommand{\corollaryname}{Corollary}

\newtheorem*{cor*}{\protect\corollaryname}

\providecommand{\propositionname}{Proposition}

\newtheorem*{prop*}{\protect\propositionname}

\providecommand{\remarkname}{Remark}
\theoremstyle{remark}

\theoremstyle{remark}
\newtheorem*{rem*}{\protect\remarkname}

\providecommand{\assumptionname}{Assumption}
\theoremstyle{definition}

\theoremstyle{definition}
\newtheorem*{assumption*}{\protect\assumptionname}

\providecommand{\claimnname}{Claim}

\newtheorem*{claim*}{\protect\claimnname}

\providecommand{\definitionname}{Definition}
\theoremstyle{definition}

\theoremstyle{definition}
\newtheorem*{defn*}{\protect\definitionname}

\providecommand{\examplename}{Example}
\theoremstyle{definition}

\theoremstyle{definition}
\newtheorem*{example*}{\protect\examplename}

%% file: root.bbl
\begin{thebibliography}{10}
\def\url#1{}
\csname url@samestyle\endcsname
\providecommand{\newblock}{\relax}
\providecommand{\bibinfo}[2]{#2}
\providecommand{\BIBentrySTDinterwordspacing}{\spaceskip=0pt\relax}
\providecommand{\BIBentryALTinterwordstretchfactor}{4}
\providecommand{\BIBentryALTinterwordspacing}{\spaceskip=\fontdimen2\font plus
\BIBentryALTinterwordstretchfactor\fontdimen3\font minus
  \fontdimen4\font\relax}
\providecommand{\BIBforeignlanguage}[2]{{%
\expandafter\ifx\csname l@#1\endcsname\relax
\typeout{** WARNING: IEEEtran.bst: No hyphenation pattern has been}%
\typeout{** loaded for the language `#1'. Using the pattern for}%
\typeout{** the default language instead.}%
\else
\language=\csname l@#1\endcsname
\fi
#2}}
\providecommand{\BIBdecl}{\relax}
\BIBdecl

\bibitem{SCC.Kalman1964}
R.~E. Kalman, ``When is a linear control system optimal?'' \emph{J. Basic
  Eng.}, vol.~86, no.~1, pp. 51--60, 1964.

\bibitem{SCC.Boyd.Ghaoui.ea1994}
S.~Boyd, L.~E. Ghaoui, E.~Feron, and V.~Balakrishnan, \emph{Linear matrix
  inequalities in system and control theory}.\hskip 1em plus 0.5em minus
  0.4em\relax SIAM, 1994.

\bibitem{SCC.Ng.Russell2000}
A.~Y. Ng and S.~Russell, ``Algorithms for inverse reinforcement learning,'' in
  \emph{Proc. Int. Conf. Mach. Learn.}\hskip 1em plus 0.5em minus 0.4em\relax
  Morgan Kaufmann, 2000, pp. 663--670.

\bibitem{SCC.Abbeel.Ng2004}
P.~Abbeel and A.~Y. Ng, ``Apprenticeship learning via inverse reinforcement
  learning,'' in \emph{Proc. Int. Conf. Mach. Learn.}, 2004.

\bibitem{SCC.Abbeel.Ng2005}
------, ``Exploration and apprenticeship learning in reinforcement learning,''
  in \emph{Proc. Int. Conf. Mach. Learn.}\hskip 1em plus 0.5em minus
  0.4em\relax ACM, 2005, pp. 1--8.

\bibitem{SCC.Ratliff.Bagnell.ea2006}
N.~D. Ratliff, J.~A. Bagnell, and M.~A. Zinkevich, ``Maximum margin planning,''
  in \emph{Proc. Int. Conf. Mach. Learn.}, 2006.

\bibitem{SCC.Ziebart.Maas.ea2008}
B.~D. Ziebart, A.~Maas, J.~A. Bagnell, and A.~K. Dey, ``Maximum entropy inverse
  reinforcement learning,'' in \emph{Proc. AAAI Conf. Artif. Intel.}, 2008, pp.
  1433--1438.

\bibitem{SCC.Jaynes1957}
E.~T. Jaynes, ``Information theory and statistical mechanics,'' \emph{Phys.
  Rev.}, vol. 106, no.~4, pp. 620--630, May 1957.

\bibitem{SCC.Ziebart.Bagnell.ea2010}
B.~D. Ziebart, J.~A. Bagnell, and A.~K. Dey, ``Modeling interaction via the
  principle of maximum causal entropy,'' in \emph{Proc. Int. Conf. Mach.
  Learn.}, Sep. 2010, pp. 1255--1262.

\bibitem{SCC.Boularias.Kober.ea2011}
A.~Boularias, J.~Kober, and J.~Peters, ``Relative entropy inverse reinforcement
  learning,'' in \emph{Proc. Int. Conf. Artif. Intell. Stat.}, G.~Gordon,
  D.~Dunson, and M.~Dud{\'{i}}k, Eds., vol.~15.\hskip 1em plus 0.5em minus
  0.4em\relax JMLR W\&CP, 2011.

\bibitem{SCC.Ramachandran.Amir2007}
\BIBentryALTinterwordspacing
D.~Ramachandran and E.~Amir, ``Bayesian inverse reinforcement learning,'' in
  \emph{Proc. Int. Joint Conf. Artif. Intell.}\hskip 1em plus 0.5em minus
  0.4em\relax San Francisco, CA, USA: Morgan Kaufmann Publishers Inc., 2007,
  pp. 2586--2591.  \url{http://dl.acm.org/citation.cfm?id=1625275.1625692}
\BIBentrySTDinterwordspacing

\bibitem{SCC.Neu.Szepesvari2007}
G.~Neu and C.~Szepesvari, ``Apprenticeship learning using inverse reinforcement
  learning and gradient methods,'' in \emph{Proc. Anu. Conf. Uncertain. Artif.
  Intell.}\hskip 1em plus 0.5em minus 0.4em\relax Corvallis, Oregon: AUAI
  Press, 2007, pp. 295--302.

\bibitem{SCC.Syed.Schapire2008}
\BIBentryALTinterwordspacing
U.~Syed and R.~E. Schapire, ``A game-theoretic approach to apprenticeship
  learning,'' in \emph{Advances in Neural Information Processing Systems 20},
  J.~C. Platt, D.~Koller, Y.~Singer, and S.~T. Roweis, Eds.\hskip 1em plus
  0.5em minus 0.4em\relax Curran Associates, Inc., 2008, pp. 1449--1456.
  \url{http://papers.nips.cc/paper/3293-a-game-theoretic-approach-to-apprenticeship-learning.pdf}
\BIBentrySTDinterwordspacing

\bibitem{SCC.Levine.Popovic.ea2010}
\BIBentryALTinterwordspacing
S.~Levine, Z.~Popovic, and V.~Koltun, ``Feature construction for inverse
  reinforcement learning,'' in \emph{Advances in Neural Information Processing
  Systems 23}, J.~D. Lafferty, C.~K.~I. Williams, J.~Shawe-Taylor, R.~S. Zemel,
  and A.~Culotta, Eds.\hskip 1em plus 0.5em minus 0.4em\relax Curran
  Associates, Inc., 2010, pp. 1342--1350.
  \url{http://papers.nips.cc/paper/3918-feature-construction-for-inverse-reinforcement-learning.pdf}
\BIBentrySTDinterwordspacing

\bibitem{SCC.Levine.Koltun2012}
\BIBentryALTinterwordspacing
S.~Levine and V.~Koltun, ``Continuous inverse optimal control with locally
  optimal examples,'' in \emph{Proc. Int. Conf. Mach. Learn.}, J.~Langford and
  J.~Pineau, Eds.\hskip 1em plus 0.5em minus 0.4em\relax New York, NY, USA:
  ACM, 2012, pp. 41--48.  \url{http://icml.cc/2012/papers/43.pdf}
\BIBentrySTDinterwordspacing

\bibitem{SCC.Levine.Popovic.ea2011}
\BIBentryALTinterwordspacing
S.~Levine, Z.~Popovic, and V.~Koltun, ``Nonlinear inverse reinforcement
  learning with gaussian processes,'' in \emph{Advances in Neural Information
  Processing Systems 24}, J.~Shawe-Taylor, R.~S. Zemel, P.~L. Bartlett,
  F.~Pereira, and K.~Q. Weinberger, Eds.\hskip 1em plus 0.5em minus 0.4em\relax
  Curran Associates, Inc., 2011, pp. 19--27.
  \url{http://papers.nips.cc/paper/4420-nonlinear-inverse-reinforcement-learning-with-gaussian-processes.pdf}
\BIBentrySTDinterwordspacing

\bibitem{SCC.Michini.How2012}
B.~Michini and J.~P. How, ``Bayesian nonparametric inverse reinforcement
  learning,'' in \emph{Machine Learning and Knowledge Discovery in Databases},
  ser. Lecture Notes in Computer Science, P.~A. Flach, T.~D. Bie, and
  N.~Cristianini, Eds.\hskip 1em plus 0.5em minus 0.4em\relax Springer Berlin
  Heidelberg, 2012, vol. 7524, pp. 148--163.

\bibitem{SCC.Mombaur.Truong.ea2010}
K.~Mombaur, A.~Truong, and J.-P. Laumond, ``From human to humanoid
  locomotion---an inverse optimal control approach,'' \emph{Auton. Robot.},
  vol.~28, no.~3, pp. 369--383, 2010.

\bibitem{SCC.Michini.Walsh.ea2015}
B.~Michini, T.~J. Walsh, A.~A. Agha-Mohammadi, and J.~P. How, ``Bayesian
  nonparametric reward learning from demonstration,'' \emph{IEEE Trans.
  Robot.}, vol.~31, no.~2, pp. 369--386, Apr. 2015.

\bibitem{SCC.Vamvoudakis.Lewis2010}
K.~Vamvoudakis and F.~Lewis, ``{Online actor-critic algorithm to solve the
  continuous-time infinite horizon optimal control problem},''
  \emph{Automatica}, vol.~46, no.~5, pp. 878--888, 2010.

\bibitem{SCC.Bian.Jiang.ea2014}
T.~Bian, Y.~Jiang, and Z.-P. Jiang, ``Adaptive dynamic programming and optimal
  control of nonlinear nonaffine systems,'' \emph{Automatica}, vol.~50, no.~10,
  pp. 2624--2632, 2014.

\bibitem{SCC.Modares.Lewis2014}
H.~Modares and F.~L. Lewis, ``Optimal tracking control of nonlinear
  partially-unknown constrained-input systems using integral reinforcement
  learning,'' \emph{Automatica}, vol.~50, no.~7, pp. 1780--1792, 2014.

\bibitem{SCC.Kamalapurkar.Walters.ea2016}
\BIBentryALTinterwordspacing
R.~Kamalapurkar, P.~Walters, and W.~E. Dixon, ``Model-based reinforcement
  learning for approximate optimal regulation,'' \emph{Automatica}, vol.~64,
  pp. 94--104, Feb. 2016.
  \url{http://www.sciencedirect.com/science/article/pii/S0005109815004392}
\BIBentrySTDinterwordspacing

\bibitem{SCC.Wang.Liu.ea2016}
D.~Wang, D.~Liu, H.~Li, B.~Luo, and H.~Ma, ``An approximate optimal control
  approach for robust stabilization of a class of discrete-time nonlinear
  systems with uncertainties,'' \emph{IEEE Trans. Syst. Man Cybern. Syst.},
  vol.~46, no.~5, pp. 713--717, 2016.

\bibitem{SCC.Kamalapurkar2017}
\BIBentryALTinterwordspacing
R.~Kamalapurkar, ``Online output-feedback parameter and state estimation for
  second order linear systems,'' in \emph{Proc. Am. Control Conf.}, Seattle,
  WA, USA, May 2017, pp. 5672--5677.
  \url{http://ieeexplore.ieee.org/document/7963838/}
\BIBentrySTDinterwordspacing

\bibitem{SCC.Ioannou.Sun1996}
P.~Ioannou and J.~Sun, \emph{Robust adaptive control}.\hskip 1em plus 0.5em
  minus 0.4em\relax Prentice Hall, 1996.

\bibitem{SCC.Xian.Queiroz.ea2004}
B.~Xian, M.~S. de~Queiroz, D.~M. Dawson, and M.~McIntyre, ``A discontinuous
  output feedback controller and velocity observer for nonlinear mechanical
  systems,'' \emph{Automatica}, vol.~40, no.~4, pp. 695--700, 2004.

\end{thebibliography}
